\documentclass[preprint,12pt]{elsarticle}

\usepackage{xcolor}
\usepackage{siunitx}
\usepackage{amssymb}
\usepackage{amsmath}
\journal{Optics \& Laser Technology}

\begin{document}

\begin{frontmatter}

\title{Fiber-based high-speed fringe projection profilometry}

\author[1]{Steven~Johnson}
\author[1]{Hal~Gee}
\author[1]{Faith~Nwachi}
\author[1]{Simon~Peter~Mekhail}
\affiliation[1]{School of Physics and Astronomy, University of Glasgow, Glasgow, G12 8QQ, United Kingdom}

\begin{abstract}
    Phase shift profilometry has become a standard technique for recording the surface of an object using the projection of sinusoidal patterns. We present a method of projecting the patterns via a pair of single-mode optical fibers mounted within a needle. The system presented can capture sub-millimetre precision depth profiles at 60 frames per second. The key features of the system that enable the high-speed imaging are the creation of an automated calibration system and developments in the image unwrapping and reconstruction algorithms to produce high fidelity surface measurements. These methods produce a compact projection system capable of accurate and fast profilometry.
\end{abstract}

\end{frontmatter}


\section{Introduction}

The use of 3D imaging to study the profile of objects has been used widely within engineering, medicine and art \cite{Geng2011, Yang2024_OverviewStructuredLight, VANDERJEUGHT2016_Review}. The ability to perform imaging through a narrow-gauge needle would enable surface imaging with minor impact on a patient; this would allow surgeons to make a better assessment of tissues before more traumatic surgical options are explored. The method proposed in this paper uses a pair of single-mode optical fibers within a narrow-gauge needle to enable fringes to be projected onto a surface via a very narrow hole and capture a profile of the area to investigate. 

Optical fibers are well known to be excellent sensors for measuring displacement and imaging applications \cite{Alayli2004_FiberDisplacement}. Often a digital projector is required to project the patterns for profilometry, but it has been previously shown that profile measurements can be performed by projecting patterns onto an object via an optical fiber \cite{Fang2001, Pennington2001_MinaturizedProfiler, Chen2014_FiberProjection}. However, these studies were measuring small-scale objects and would be unsuitable for measurements of macroscopic objects. There has been other work using additional optical fibers to deliver the light, such as with a multi-core fiber \cite{Bulut2005_4corefibre}, or by using an apparatus where a Lloyd's mirror is used at the end of an optical fiber \cite{Kosoglu2016_LloydsMirror}. An alternative technology for imaging large objects at a distance has been shown to be possible via imaging through a multi-mode fiber using the optical modes of the fiber and a pulsed laser to perform time-of-flight measurements to gather the 3D information. This however required costly high speed detectors, spatial phase shaping equipment, and was limited in frame rate due to the scanning nature of the method used \cite{Stellinga2021}. In this work we show a higher spatial, axial, and temporal resolution with a comparatively simple and cost effective setup.

\textcolor{black}{There are a range of methods that use the projection of fringes to perform profilometry. In this paper we use phase shift profilometry (PSP) \cite{Cai2016} where multiple images are captured with varying the phase of the projected fringe patterns. Some methods such as Fourier transformation profilometry \cite{Su_2001_FTP} use a single frame of fixed a fringe pattern projected onto the object. A method known as modulation measuring profilometry \cite {LU_2016_MMP} uses a crossed grating and the where the vertical and horizontal components are separated in image space. Another technique uses the Moir\'e principle, where a grating is placed in front of the light source and the camera, these overlapping gratings are used to do Moir\'e profilometry \cite{Li_2017_Moire}, gratings can also be created through digital projection \cite{Li2019_Moire}.}

The emission of coherent light from two single-mode fibers produces the classical ``Young's slits'' sinusoidal interference pattern. Adjusting the phase of one of the outputs will move the the position of the optical fringes on the object, this shifting of the phase upon the object enables depth imaging to be performed via PSP, a well know technique for the reconstruction of surface profiles of objects. The technique is well known for its high resolution and accuracy, and has proven useful in high speed and real time 3D shape measurement \cite{Pages2003, Xu2020}. PSP uses multiple images of the fringes, at least three projected onto the object, to calculate a phase distribution and hence realise a 3D profile of a scene. Each of the fringe images consists of a different phase shift, and by measuring the different intensities at a given point across the images, the relative depth of that point can be ultimately calculated. The initial output of the measurement is a wrapped phase image, but from this wrapped phase image there is significant processing needed to gain the depth profile. \textcolor{black}{There have been developments in unwrapping methods, such as a careful choice of the guided path contours \cite{Estrada_2012_Unwrap}, and rejecting outlier values to reduce confusion within the unwrapping \cite{RIVERA_2015_Unwrap}. Alternative unwrapping methods would be a least squares approach \cite{Juarez_2014_leastSquares, GUO_2014_LeastSquares}, curtain-type phase unwrapping algorithm \cite{Xu_2022_CurtainType} or using a ternary complementary Gray code phase unwrapping method \cite{Zheng_2017_GrayCode, Wei_2024_Graycode}. One effective method that has produced promising results within unwrapping is deep learning \cite{Wang2019_DL, HUANG2022_DLunwrap}, although this method requires large training sets to produce accurate results.}
We require an unwrapping algorithm where an appropriate path is chosen to reconstruct the 2D depth image in an efficient way. 

In this paper we present a needle-based optical fiber projection system, which can be used in minimally invasive profile scans of an object. Imaging of surfaces has been demonstrated at 60 frames per second with a robust reconstruction approach using a simple optical set up, alongside which an automatic phase calibration routine has been developed. The method by which the calculation of an accurate 3D profile is measured has been improved upon from previously published works. We have also introduced novel image processing which leads to an increase in speed of reconstruction. 

\section{Methodology}

\subsection{Experimental set-up}

\begin{figure}
    \centering
    \includegraphics[width=0.95\textwidth]{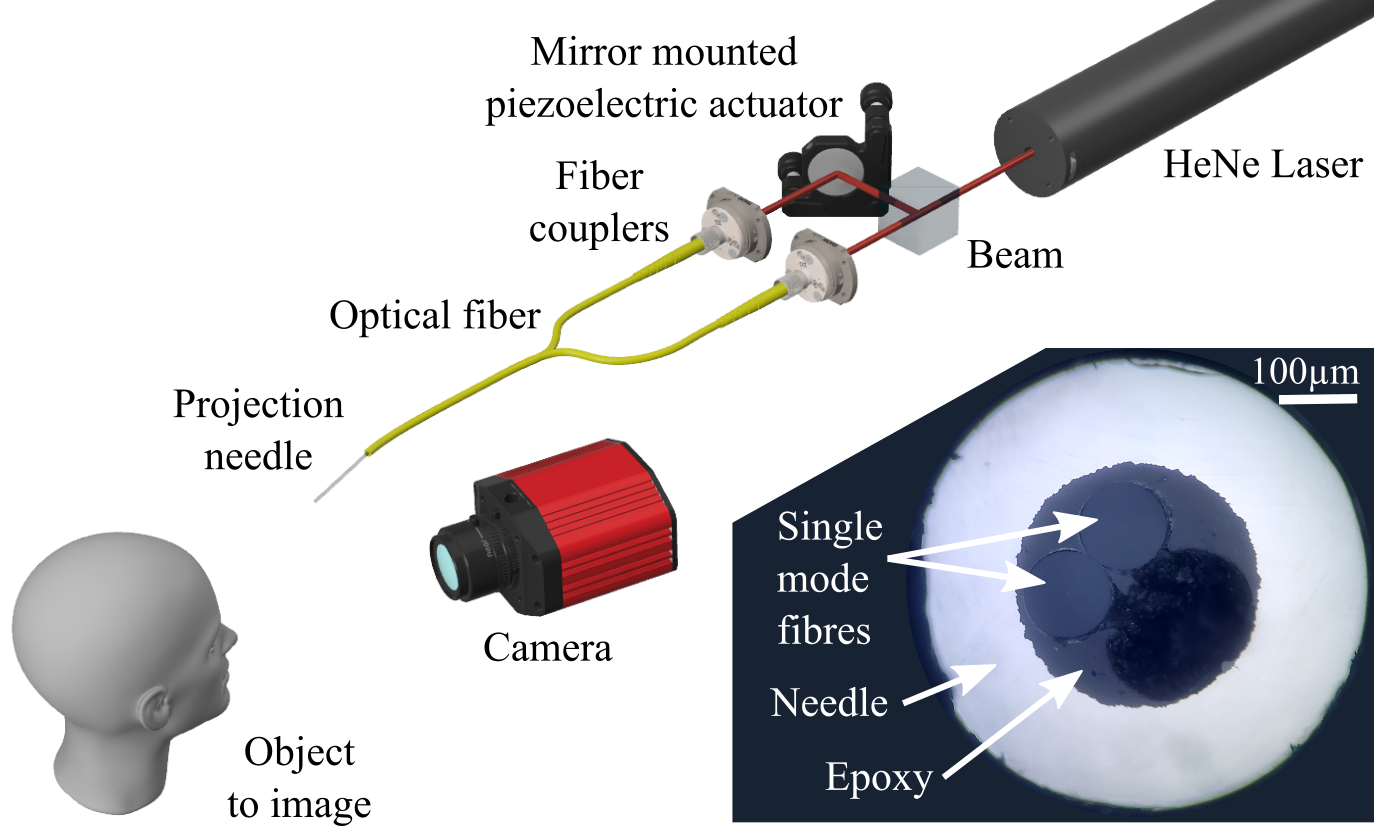}
    \caption{The experimental layout. A HeNe laser is split with a 50:50 beam splitter and aligned into fiber couplers with two mirrors per beam. One mirror has a piezoelectric actuator mounted behind it to enable its angle to be adjusted. The actuator is controlled with a high-voltage signal amplified from a DAC connected to the computer. The optical fiber assembly has two fibers within a 23 gauge needle, from which light is projected onto the object. An image of each fringe pattern is taken by a camera, which triggers a digital-to-analog converter (DAC) for synchronisation of pattern projection and image capture. The inset image shows the polished needle face with two single-mode optical fibers fixed inside, photographed through a telescope The scale bar is \SI{100}{\micro\metre}.}
    \label{fig:Layout}
\end{figure}

The experimental set up is shown in figure \ref{fig:Layout}. Profilometry was performed using two interfering light fields transmitted through single mode optical fibers mounted in a needle. Coherent light from a Helium-Neon laser was coupled into each of the optical fibers, after which the light interferes at the object to produce fringes varying along the axis of separation of the optical fibers at a plane normal to the needle.
The optical set-up used a minimal number of components to split the coherent light source and couple it into two optical fibers: a HeNe laser (Thorlabs HNL050LB) split with a 50:50 beam splitter. The two beams were coupled into either side of the optical fiber assembly via steering mirrors. To adjust the eventual fringe position, the optical path length of one arm was adjusted by shifting the position of one of the mirrors with a piezoelectric actuator (PK4FA2P1) driven by a high voltage power supply (Piezomechanick SQV 1/1000). The piezoelectric actuator was mounted in a custom-designed 3D-printed mirror mount. The voltage was controlled via an I\textsuperscript{2}C DAC (AD5659R) driven by an Arduino UNO-equivalent micro-controller board. \textcolor{black}{Changing the voltage to the actuator will change its length. One side of the mirror rests on the actuator and will shift in position when adjusted, this would change the length of light path and result in a difference in phase in the interference pattern of the light.}  The light from the fibers was projected onto the object and imaged with a camera (Alvium 1800 U-240C, 2.4 MP), onto which a narrow band-pass filter (\SI{635}{nm}) was fitted to filter background light from the scene. The camera drove a digital control line connected to the micro-controller in order to synchronise image capture with phase change, whereby every falling edge of the camera exposure signal triggered the change in actuator control voltage and hence moved the fringes.

\subsection{Fiber assembly construction}
The needle used for imaging was developed for the specific use of fringe projection. The light emerging from the fibers was projected directly onto the target. Single-mode fiber was used for this project as opposed to multi-mode fiber because of the comparatively small diameter and hence core-to-core distance achievable. This distance between the single-mode optical fibers plays a crucial role in the scale of the patterns produced. Furthermore, multimode fibers would require more control over the input field to ensure modal dispersion was accounted for.

The optical fiber assembly consisted of two single-mode fibers fixed within a needle. To construct the fiber assembly, single-mode optical fiber (P1-630-FC) was cut in half and the protective PVC jacket was removed, followed by the layer of Kevlar strands from around the end of the fiber. The optical fiber was then stripped of its plastic coating using strippers and acetone, exposing the \SI{125}{\micro\metre} glass cladding. The stripped ends of the two optical fibers were carefully inserted into one jacket for protection. Epoxy was drawn into the 23 gauge needle (\SI{337}{\micro\metre} inner diameter and \SI{642}{\micro\metre} outer diameter) using a syringe, the two fiber ends were then inserted into the epoxy-filled needle. After curing the epoxy, the end of the needle was sawed off to expose the fiber surfaces, which were polished as one using increasing grades of fiber polishing/lapping films. \textcolor{black}{The optical fibers had a numerical aperture of 0.14, this produces a cone of light emitting from the fibers that is \ang{16} across.}

\subsection{Calibration of fringe phase shifts}
An often cited problem with optical fibers relating to phase measurement is the variation in the transmission of a reliable phase due to temperature changes or strain within the fiber. With a sufficiently fast calibration scheme, both calibration and imaging can be performed before these issues become significant. The goal of calibration in our set-up is to find three voltages that can be applied cyclically to the actuator for an equal phase-shift of $2\pi/3$ between consecutive fringe projections.

Due to the hysteresis of the actuator, it is not sufficient to perform a voltage sweep and simply sample into the observed phase response at $2\pi/$3 intervals. The values obtained by doing so will not produce the same phase shifts when applied in a cycle. We begin by choosing an arbitrary value for the first cycle voltage ($V_0$), which does not vary during calibration ($\phi_0 = 0$ always, as phases are taken relative to the first image in a cycle). A range is chosen for the $\phi_1 = 2\pi/3$ ($V_1$) and $\phi_2 = 4\pi/3$ ($V_2$) voltage values, effectively bounding a rectangle in $(V_1, V_2)$-space. The calibration process is a discrete sweep of this rectangle \textcolor{black}{over a 10 by 10 uniform grid}. At each point the evenness of the measured phases is evaluated by calculating the squared error (PSE) as:

\begin{equation}
    \mbox{PSE} =  \sum^2_{n=0} \left(V_n - V_0 - \frac{2\pi}{3}n\right)^2 
\end{equation}

The cycle is found with minimum PSE, and the search ranges for $V_1$ and $V_2$ are halved and shifted such that the calculated minimum lies in the centre of the new rectangle they bound. Two more iterations of this procedure are performed, each time examining a smaller range of voltages at higher resolution, finally yielding a cycle of 3 voltages with stable, evenly spaced phase responses. \textcolor{black}{The calibration process found optimal values for the piezo actuator but also had the effect of preconditioning the system to ensure reliable phase stepping. Once this was performed imaging could proceed ensuring the fibres were not significantly disturbed. Ambient temperature changes and vibrations seemed to have negligible effects on imaging quality}.

\subsection{Profile reconstruction method}

Using three different phase patterns the phase profile can be reconstructed. As the fringes are periodic, this phase profile can only be reconstructed within the range $[-\pi, \pi)$ and is therefore known as the ``wrapped'' phase. Unwrapping this phase profile to full depth range is an important step in phase-shifting profilometry. An accurate method of unwrapping the phase to determine the shape of a plane requires a computational approach that evaluates the quality of the data at each point and guides the unwrapping path, \textcolor{black}{known as the quality guided path}. Due to the ambiguous nature of the wrapped data, it is possible to reconstruct sections of the target object in the wrong position depending on the path used. Analysis in the frequency domain can be used to make the unwrapping method more robust to the noise with the image \cite{Ghiglia1994}. Further improvements can be made using phase unwrapping based upon the transport of intensity, where the gradient of the intensity is used \cite{Martinez-Carranza2017}\textcolor{black}{, which was the method implemented here}. Accessing the quality of the data at each point is achieved using windowed Fourier filtering (WFF), this produces a quality map that is used to choose the path over which the data is unwrapped \cite{Kemao2007,Kemao2008}. To speed up the reconstruction a stepping was used, this produced equivalent results to the method of performing the WFF at each point within the area over which the phase unwrapping is performed.

Individual images of a target under fringe pattern illumination will contain the pattern intensity phase-modulated in proportion to target depth. The fringe patterns are projected with a difference in global phase shift of $2\pi/3$ between each, for a total of three equally phase-spaced images:
\begin{eqnarray}
    I_1(x,y) &=& I_\textrm{bg} (x,y) + I_\textrm{amp} (x,y) \cos\left(\phi(x,y)\right),\\
    I_2(x,y) &=& I_\textrm{bg} (x,y) + I_\textrm{amp} (x,y) \cos\left(\phi(x,y) + 2\pi/3\right),\\
    I_3(x,y) &=& I_\textrm{bg} (x,y) + I_\textrm{amp} (x,y) \cos\left(\phi(x,y) + 4\pi/3\right),
\end{eqnarray}
where $I_1(x,y)$, $I_2(x,y)$, and $I_3(x,y)$ are the captured images of the illuminated target, $I_\textrm{amp} (x,y)$ is the fringe amplitude map, $I_\textrm{bg} (x,y)$ is the background intensity, and $\phi(x,y)$ is the pattern phase map without any global shifts. The phase information can be retrieved to produce the wrapped phase image $\phi'$ as follows:

\begin{equation}
    \phi' = \arctan\left[ \sqrt{3} \frac{I_1(x,y) - I_3(x,y)}{2 I_2(x,y) - I_1(x,y) - I_3(x,y)}   \right] .
    \label{eq:arctan}
\end{equation}

\begin{figure}
    \centering
    \includegraphics[width=1.0\textwidth]{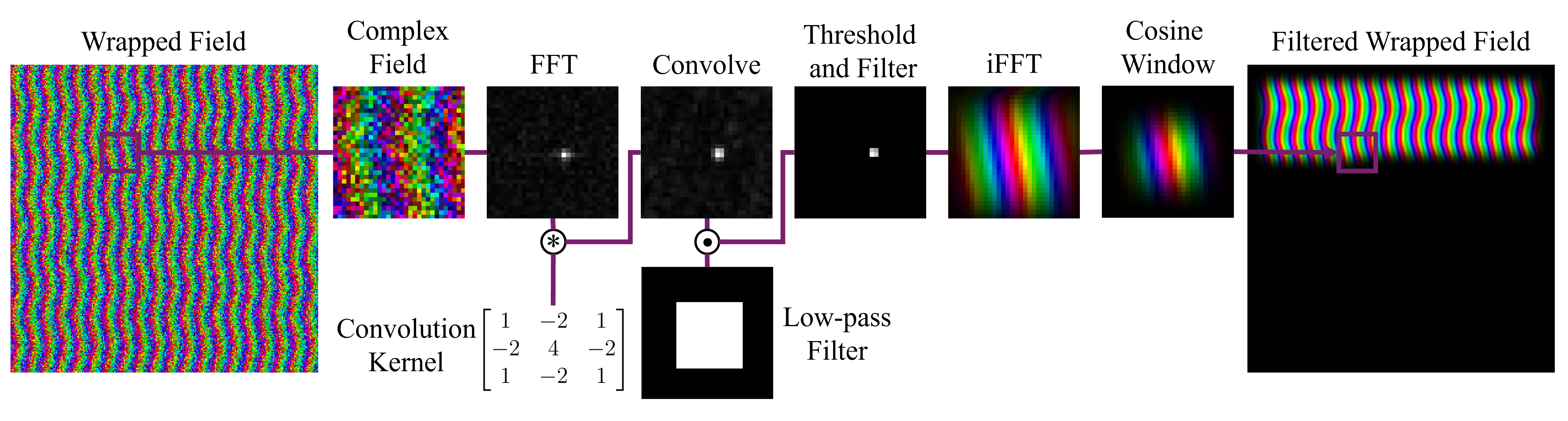}
    \caption{Windowed Fourier filtering (WFF) was used to efficiently denoise the wrapped phase image. A window of the complex field image was Fourier transformed, convolved with a convolution kernel, and a low-pass filter and thresholding applied. This field was then inverse Fourier transformed and a cosine window applied, before being added to a new filtered image.}
    \label{fig:WFF}
\end{figure}

Phase unwrapping is the process that converts the wrapped phase to the absolute phase. Windowed Fourier filtering not only filters the wrapped phase image, being able to recover usable phase data under very low contrast-to-noise ratios, but also provides a map of data ``quality'' with almost no extra computation. The quality is, in effect, a measure of local similarity to an ideal wrapped phase ramp around a given point, and can be used to guide the unwrapping process preferentially over areas least likely to introduce phase errors. Previous methods were not able to perform fast production of a surface from the acquired images.
The WFF method used was based upon the methods implemented in papers by Kemao et al. \cite{Kemao2007,Kemao2008}, with novel additions in windowing and parallelisation. The method is details in figure \ref{fig:WFF} and the method runs as follows:

\begin{enumerate}
\item Choose a window size $w$, window stride $s$ \textcolor{black}{(where $w>s$), and a Fourier coefficient threshold.}
\item From the wrapped phase image, take subsections of size $w\times w$, whose top left corners lie on a grid with pitch equal to the stride $s$. These subsections will be called ``windows''.
\item Apply a 2-dimensional fast Fourier transform (FFT) to \textcolor{black}{complex representation of} each window.
\item In each transformed window, \textcolor{black}{perform a periodic boundary convolution by replacing} every value with the finite difference calculated between it and the surrounding 8 values, according to the kernel matrix:
\[
\begin{bmatrix}
1  & -2 &  1\\
-2 &  4 & -2\\
1  & -2 &  1\\
\end{bmatrix}
\]
\textcolor{black}{This kernel is akin to element-wise multiplication of the window with a 2-dimensional cosine function with a horizontal and vertical spatial period $w$ in pixel space.}
\item Apply a \textcolor{black}{square-shaped top-hat low-pass filter with width $w/2 \times w/2$ concentric with the window} to each window to remove the higher frequency components, and set remaining coefficients with magnitude below the Fourier coefficient threshold to zero.
\item Apply a 2-dimensional inverse FFT to each window.
\item Multiply the windows by a cosine function, defined as
\begin{equation}
W(i,j) =\left(1-\cos\left(\frac{2\pi i}{w}\right)\right) \left(1-\cos\left(\frac{2\pi j}{w}\right)\right)    
\end{equation}
This multiplication (up to a constant factor) corresponds exactly to the finite difference operation above. \textcolor{black}{This step also has the benefit of suppressing ringing from Fourier filtering especially in cases where $s>1$.}
\item Sum all of the windows into a new initially-zero image of the same size as the original, each window being added onto the region from which it was taken.
\item Divide each pixel in the new image by the number of windows which landed on it during addition. For an image coordinate $(i, j)$, this number is denoted $N$ and is given by:
\begin{equation}
n(i)=\left\lfloor \frac{i}{s} \right\rfloor - \left\lfloor \frac{i-w}{s} \right\rfloor
\end{equation}
\begin{equation} 
N(i,j) = n(i) \cdot n(j)
\end{equation}
\end{enumerate}

While the finite difference operation is itself not more efficient than multiplication by a pre-computed cosine function, the fact that it is performed after the Fourier transform allows for far less computation than in previous methods. This is because each window no longer needs entirely independent Fourier transformation - one may take a subsection of the image with full width but height equal to the window size, and first perform only a column-wise (vertical) FFT. Then each window along this subsection requires only an additional row-wise (horizontal) FFT. This is not possible if each window must be multiplied by an arbitrary function before a 2D FFT. Further optimisation may be available; the FFTs of overlapping windows are related, and can be calculated as a ``running'' FFT as is common in real-time spectrogram calculation. However, this is complicated by the variable stride, which we typically set greater than 1, and tests gave no major speed improvements with this optimisation scheme.

The splitting of the image into rows of windows as just described also opens an easy door into parallelisation. Each row of windows may be calculated independently and summed to produce a horizontal portion of the output image. Many producers can thus generate portions of the filtered output, while a single consumer collects and sums them into the final output image. This is the parallelisation scheme implemented.

Lastly, the use of a stride greater than 1 allows for much faster processing with negligible ill effect, as long as it is small compared to the window size (typically $\leq 25\%$). The number of windows processed is proportional to the inverse square of the stride, thus even small strides significantly reduce computation. For the results presented a stride of 4 is used, and a window size for applying the FFT of 32 is used.

The result of the above algorithm is a complex-valued image. Its argument is the filtered wrapped phase map, and its magnitude is a measure of how much spatial frequency information passed through filtering unaffected in the neighbourhood of each pixel. The latter serves well as a map of image ``quality'', as areas in the original image with little fringe information will be filtered to low magnitude. The filtered wrapped phase map can then be unwrapped via a path through the image that begin at high quality and descend, minimising the effect on the global solution of errors in low-quality regions \cite{Zhao2011}. In the particular implementation chosen, the path is constructed as unwrapping proceeds via the quality guided path \cite{Herraez2005}. This unwrapping implementation was chosen due to the minimal computational time required for this method. The list of pixels adjacent to those already unwrapped is stored as a binary max heap \cite{Firas2005}, giving efficient retrieval of the highest-quality pixel and insertion of those to be unwrapped.

\section{Experiments} 
\begin{figure}
    \centering
    \includegraphics[width=1.0\textwidth]{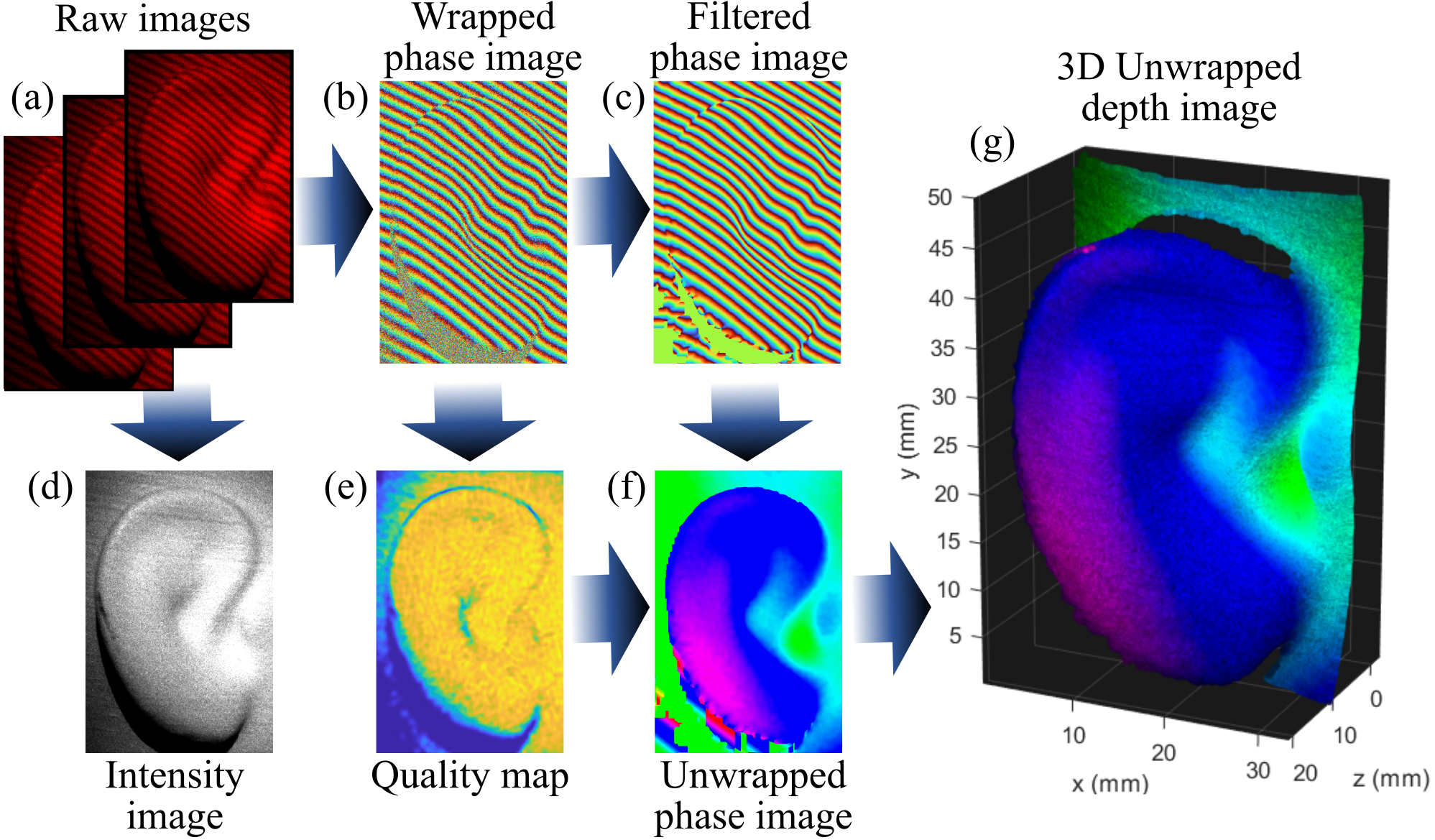}
    \caption{The ear of a mannequin is used to demonstrate the image processing steps within our system (a) The raw phase-shifted images were acquired by the camera. (b) The wrapped phase was calculated from these images to give a noisy wrapped image where the values vary between $-\pi$ and $\pi$. (c) Windowed Fourier filtering (WFF) was used to filter the image and also calculate the quality map. (d) The intensity image was calculate by summing together the three phase-shifted images. (e) The quality map was calculated from the WFF, and will be used to guide the unwrapping of the image via the quality guided path method. (f) The unwrapped image was produced from the filtered phase image to give the 3D depth in terms of phase. (g) The depth was calculated from the unwrapped phase image and the known geometry of the projection system. The point cloud produced has the intensity given from the intensity image and points removed where there is low quality as shown in the quality map.}
    \label{fig:EarPlots}
\end{figure}

Fringe projection profilometry was performed on objects approximately \SI{200}{\milli\metre} from the camera and fibers, \textcolor{black}{an area \SI{60}{\milli\metre} across was illuminated with fringes}. The stages of reconstruction are shown in figure \ref{fig:EarPlots}, where three raw images were taken under fringe projection with a $2\pi/3$ phase difference between each. From the raw images, the wrapped phase image was calculated using equation \ref{eq:arctan}. The wrapped image had significant error due to the propagation of speckle noise through the phase calculation. This was especially the case where the contrast-to-noise ratio was low due to camera readout noise. 
\textcolor{black}{The speckle noise was visible in the camera images but was spatially a high frequency component that was removed from the image with a low-pass filter, which did not have a major impact on the eventual profile measurements.}
WFF was used to calculate the quality map and filtered phase image. The unwrapping function used a quality-guided path as calculated from the quality map to accurately unwrap the filtered phase image. This produced the unwrapped phase image. To convert between phase and depth of a target point, we use the relationship between the fringe size on the object and the ratio of the distance to the object to the distance between the camera and the projection needle. For the final 3D unwrapped depth image, the quality map is used to threshold which points are plotted, such that points with very low quality are not displayed. The intensity image is used as the brightness of each point.

\begin{figure}
    \centering
    \includegraphics[width=1.0\textwidth]{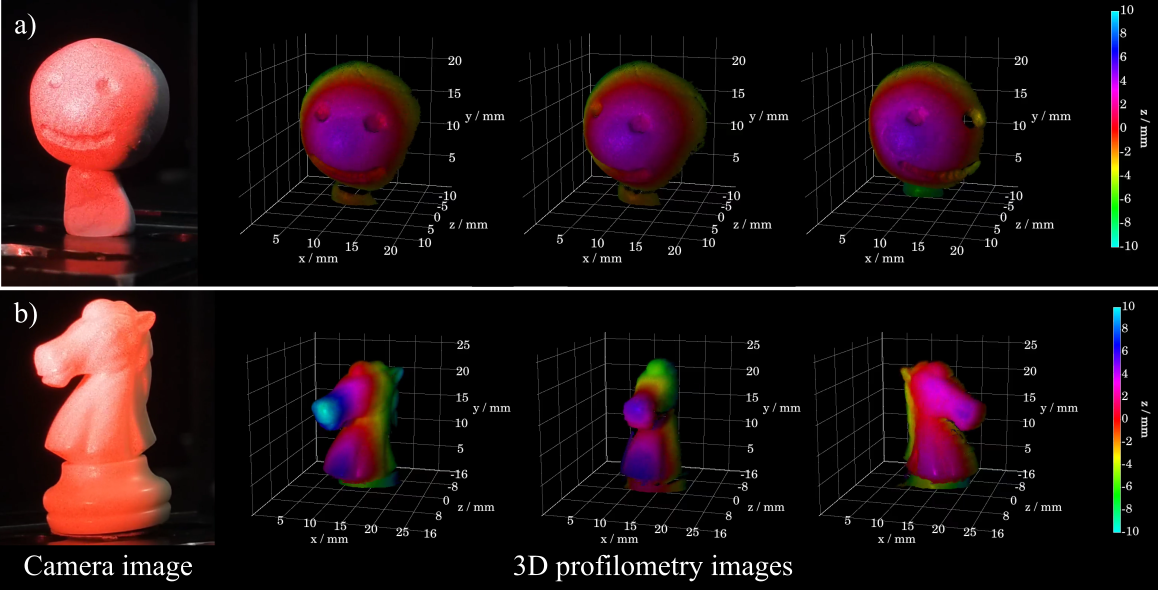}
    \caption{A camera image and three profilometry images taken from different angles for the object under measurement. a) A head made from mouldable putty. b) A knight chess piece. The object was rotated and image data recorded at 60 frames per second. Videos of the acquisition can be found in the supplementary materials.}
    \label{fig:Results}
\end{figure}

To demonstrate the profilometry quality of the system several objects were imaged and its profile measured. The objects were mounted on a rotating stand and rotated at a constant rate during acquisition. The profiles measured for two of these objects is shown in figure \ref{fig:Results}, where videos of the moving object and the profile measurements are shown in the supplementary materials. The digital control line between the camera and the phase-stepping enabled the acquisition of raw phase images to be performed at 60 frames per second. The videos within the supplementary materials produced from an overlapping of three input frames, such that a sequential triplet of frames are used to calculate the profile. One limit to the acquisition speed was the exposure time per frame, which had to be long enough to yield sufficient contrast-to-noise that the extracted phase map would be usable. This time would be reduced under brighter illumination. \textcolor{black}{In comparison to other high-speed projection profilometry systems \cite{VANDERJEUGHT2016_Review} 60 frames per second is a competitive frame rate, although this time does not include the reconstruction time for the calculation of the profile in our system.}

Within the acquisitions, jumps of the profile position could be seen due to the analysis incorrectly calculating the depth position. This would occur where there was a step change in the surface height, where the quality guided path could not link the areas to the previous surface, these effect are also more likely to occur at the edges of the illuminated area or at the side of an object where low amounts of the scattered light are not returned to the camera. 

\textcolor{black}{To characterise the system a known shape was 3D printed and compared to the profilometry measurement. A sine wave shape with an amplitude of \qty{5}{mm} and a peak spacing of \qty{16}{mm} was used to measure the profile, the global profile comparison to the known sine wave deviated due to the geometric optics of the system. However, the deviation from a fitted sine wave had a mean squared error of \qty{0.2}{mm} in the optimal needle and camera position.}

\section{Conclusion}

We have demonstrated a profilometry system that can acquire images at a rate of 60 frames per second, producing sub-millimetre accuracy of the reconstructed surface. The system uses a custom-developed needle with two internally fixed single-mode optical fibers. Such a system could have applications in medical or surveying purposes. Within the development of the project, an automated calibration scheme has been developed to control the phase offsets used in fringe projection. This calibration process achieved accurate and repeatable phase shifting, and the calibration was straightforward to perform. We also present improvements in the speed of image processing with a wider stride in the WFF, reducing the number of Fourier transforms required, and windowing horizontal sections of the image in parallel. 

\section{Acknowledgments}
Engineering and Physical Sciences Research Council (EP/T00097X/1,  EP/T517896/1). The Leverhulme Trust Early Career Fellowships. We thank Miles Padgett for useful discussions during the development of the project. 

\section{Author Contributions}
S.J. conceived the project through discussion with S.P.M. The fiber assembly was constructed by F.N. along with the initial experimental set up. The WFF methodology and writing of the analysis code was developed by H.G. with support from the others. The initial manuscript was written by S.J. All authors reviewed the manuscript.


\section{Disclosures}
The authors declare no competing interests. 

\bibliographystyle{elsarticle-num}
\bibliography{bibfile}

\end{document}